\documentclass[twocolumn,secnumarabic,amssymb, nobibnotes, aps, prd, superscriptaddress]{revtex4-1}
\usepackage[T1]{fontenc}
\usepackage{graphicx}
\usepackage{color}
\usepackage{amsmath}
\usepackage{upgreek}
\usepackage{gensymb}
\usepackage[left,modulo]{lineno}
\usepackage{upgreek}
\begin{document}
\title{Temperature-Controlled Slip of Polymer Melts on Ideal Substrates}
%\title{A comparison between the temperature dependence of the slip length in polymer melts to the polymer-surface friction activation energies}
%\title{Link between temperature dependence of PDMS polymer melt slip and elastomeric friction}
\author{Marceau H\'enot}
\affiliation{Laboratoire de Physique des Solides, CNRS, Univ. Paris-Sud, Universit\'e
Paris-Saclay, 91405 Orsay Cedex, France}
\author{Marion Grzelka}
\affiliation{Laboratoire de Physique des Solides, CNRS, Univ. Paris-Sud, Universit\'e
Paris-Saclay, 91405 Orsay Cedex, France}
\author{Jian Zhang}
\affiliation{Laboratoire de Physique des Solides, CNRS, Univ. Paris-Sud, Universit\'e
Paris-Saclay, 91405 Orsay Cedex, France}
\author{Sandrine Mariot}
\affiliation{Laboratoire de Physique des Solides, CNRS, Univ. Paris-Sud, Universit\'e
Paris-Saclay, 91405 Orsay Cedex, France}
\author{Iurii Antoniuk}
\affiliation{Univ Lyon, Universit\'e Lyon 1, CNRS, Ing\'enierie des Mat\'eriaux Polym\`eres, UMR 5223, F-69003, Lyon, France}
\author{Eric Drockenmuller}
\affiliation{Univ Lyon, Universit\'e Lyon 1, CNRS, Ing\'enierie des Mat\'eriaux Polym\`eres, UMR 5223, F-69003, Lyon, France}
\author{Liliane L\'eger}
\affiliation{Laboratoire de Physique des Solides, CNRS, Univ. Paris-Sud, Universit\'e
Paris-Saclay, 91405 Orsay Cedex, France}
\author{Fr\'ed\'eric Restagno}
\email[Corresponding author: ]{frederic.restagno@u-psud.fr}
\affiliation{Laboratoire de Physique des Solides, CNRS, Univ. Paris-Sud, Universit\'e
Paris-Saclay, 91405 Orsay Cedex, France}
\date{\today}
\begin{abstract}
The temperature dependence of the hydrodynamic boundary condition between a PDMS melt and two different non-attractive surfaces made of either an OTS (octadecyltrichlorosilane) self-assembled monolayer (SAM) or a grafted layer of short PDMS chains has been characterized. A slip length proportional to the fluid viscosity is observed on both surfaces. The slip temperature dependence is deeply influenced by the surfaces. The viscous stress exerted by the polymer liquid on the surface is observed to follow exactly the same temperature dependences as the friction stress of a cross-linked elastomer sliding on the same surfaces. \textcolor{black}{Far above the glass transition temperature,} these observations are rationalized in the framework of a molecular model based on activation energies: increase or decrease of the slip length with increasing temperatures can be observed depending on how the activation energy of the bulk viscosity compares to that of the interfacial Navier's friction coefficient.
\end{abstract}
\maketitle
\begin{figure}[htbp]
  \centering
  \includegraphics[width=8.5cm]{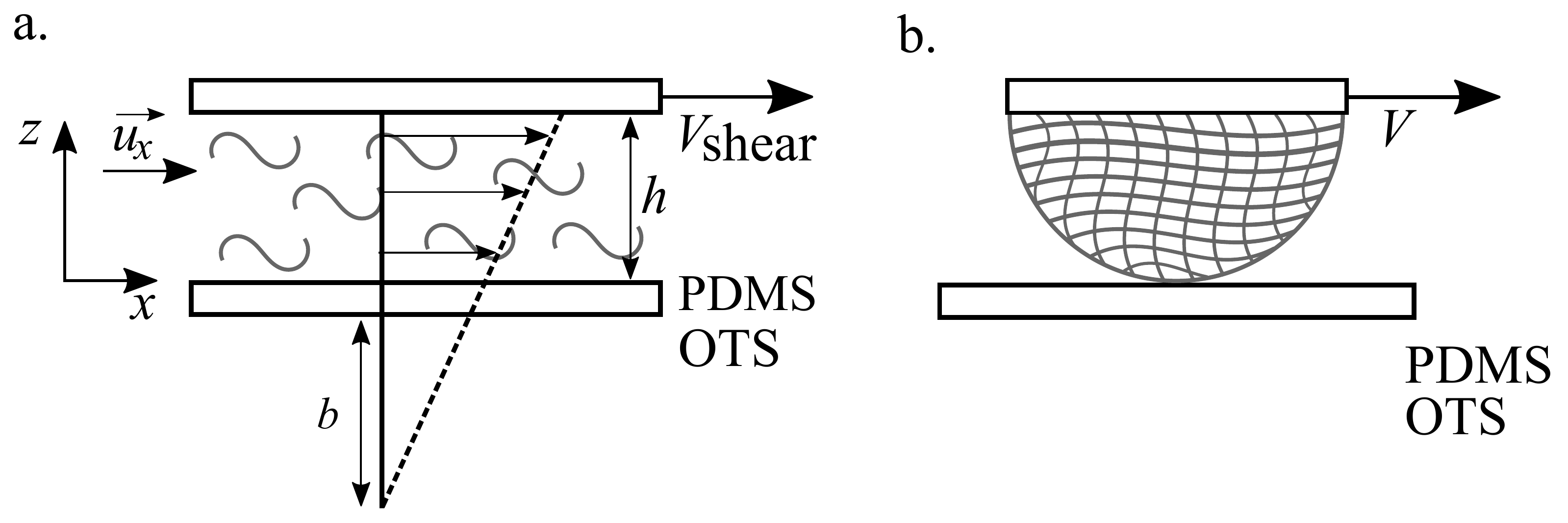}
 \caption{a. Measurement of the slip length $b$ of a PDMS melt on different surfaces by velocimetry using photobleaching. b. Measurement of the coefficient of friction $k$ of a cross-linked PDMS lens on different surfaces.}
  \label{schema}
\end{figure}

Modeling fluid flows in channels is a general problem in science and engineering. For ideal liquids, the situation is simple: there is no dissipation due to fluid movement. For real liquids, some energy is lost. Navier~\cite{navier} identified two possible sources of dissipation: bulk dissipation, associated to the friction between layers of liquid, and surface dissipation, associated to the friction of the last layer of liquid molecules sliding on the solid surface. The bulk dissipation can be obtained assuming a linear relation between the shear stress and the velocity gradient, which, for incompressible fluids, gives the Navier-Stokes equation. For surface dissipation, a classical assumption is that a liquid element adjacent to the surface assumes the velocity of the surface, i.e. a non-slip boundary condition, which leads to no surface dissipation.  Indeed, Navier, postulated the existence of a slip velocity at the surface. He proposed a linear relation between the shear stress at the solid-liquid interface and the slip velocity: $\sigma_\mathrm{fluid\rightarrow surface}=k V$, where $k$ is the interfacial friction coefficient, sometimes called the Navier's coefficient, assumed to be independent of the shear rate, and $V$ is the slip velocity. It is thus possible to define the slip length as the distance from the solid surface where the fluid velocity profile extrapolates linearly to zero (see Figure~\ref{schema}a). Balancing the viscous stress exerted by the fluid on the solid $\sigma = \eta \dot{\gamma}$, where $\eta$ is the fluid viscosity and $\dot{\gamma}$ is the shear rate, to the friction stress proposed by Navier gives:
\begin{equation}\label{equ_b}
b=\frac{\eta}{k}
\end{equation}
The slip length, if it exists, is thus the ratio of two quantities characterizing respectively bulk and surface dissipation mechanisms. In this equation, both $\eta$ and $k$ should depend on the temperature.

 Slip length determination in the case of simple fluids has been the subject of intensive experimental~\cite{chan_drainage_1985,pit_2000,cottin-bizonne_boundary_2005,schmatko_2005,neto_boundary_2005,lauga_2007} and theoretical/numerical \cite{bocquet_hydrodynamic_1994,thompson_general_1997,priezjev_molecular_2004, bocquet_flow_2007,cross2018wall} research over the last 20 years. Despite this strong activity, there is still no quantitative agreement between experiments and numerical simulations, due in part to the experimental difficulties in accurately measuring slip lengths of molecular sizes, and also to the extreme sensitivity of slip lengths to tiny molecular details of both surface and fluid. \textcolor{black}{There is at present a consensus to say that the interfacial friction of simple fluids depends on the molecular nature of the surfaces, the liquid-solid interaction energies, the local liquid ordering at surfaces and the roughness of the surfaces. However,  the effect of temperature on the slip of simple liquids remains largely unknown despite the fact that the temperature dependence constitutes an interesting way to identify the molecular mechanisms of friction at the solid-liquid interface}

    Contrary to simple liquids, polymer melts can present slip lengths much larger than the size of the molecules. This has been first inferred from the study of extrusion instabilities~\cite{petrie_instabilities_1976,elKissi_1990} and has applications in polymer extrusion~\cite{ramamurthy_1986, piau1994measurement, denn2001extrusion}, adhesion~\cite{Newby_1995, newby_1997} or lubrication in industrial processes~\cite{cayer2008drainage}. In 1979, de Gennes proposed a simple physical picture for the slip of polymer melts on ideal non-adsorbing surfaces~\cite{de_gennes_1979}. He assumed that the interfacial friction coefficient of a simple fluid made of monomers is the same as that of a polymer melt made of the same monomers. The intuitive physical argument is that in both cases monomers are sliding on the surface. The slip length would then be controlled by the fluid viscosity: $b = \eta/k_\mathrm{melt}$ with $k_\mathrm{melt}$ independent of the polymer molecular weight, and depending only on the chemical nature of the liquid and of the surface. The linear relation between slip length and viscosity has been well established by different groups~\cite{wang_molecular_1997,baumchen_reduced_2009,baumchen_slippage_2012,henot_acs_2018}.  From a theoretical point of view, using polymer fluids to investigate slip is timely since it provides an efficient tool to test the molecular mechanisms of slip: indeed, polymers enhance the degree of slip, making measurements easier, and allow one to vary the fluid viscosity in a wide range, without affecting the local interactions at the solid fluid interface. Quite recently, H\'enot {\em et al.} ~\cite{henot_acs_2018} showed experimentally that the friction stress exerted by a PDMS melt flowing on a solid surface is equal to the friction stress of a cross-linked PDMS sliding on the same solid surface, providing a simple experimental proof of the local origin of the interfacial friction.
    
    However, little is known about the effect of temperature on the slip of polymer melts. Numerical simulations report \textcolor{black}{non-monotonic} behaviors of the slip length with respect to the temperature~\cite{priezjev_molecular_2004, servantie2008temperature}. From an experimental point of view, Wang and Dra measured a slip length of HDPE almost independent of the temperature~\cite{Drda_1995}. \textcolor{black}{More recently, B\"aumchen~\textit{et al.}~\cite{baumchen_sliding_2010} compared the slip length of polystyrene on three substrates for different temperatures : they showed that the slip length can decrease or be constant with the temperature depending on the substrate.}This could be indicative of an effect of the temperature on the interfacial friction coefficient $k$.  In view of the available literature, the temperature dependence of the slip of polymer fluids appears puzzling and seems to depend on the studied system. 

  In this letter, we present an investigation of the effect of temperature on the slip length for a PDMS melt flowing on two ideal non-adsorbing surfaces: a self-assembled monolayer (SAM) of octadecyltrichlorosilane (OTS) and a layer of end-tethered non-entangled PDMS chains. The choice of these model polymer and surfaces allows us to clearly establish that either strong or small effects of the temperature on the slip length can be observed depending on the nature of the surface. Comparing the friction stresses exerted by a PDMS melt or by a cross-linked PDMS elastomer at different temperatures on the same solid, we show that it is indeed possible to separate the effect of temperature on respectively bulk and surface dissipation. This allows us to propose what we think to be a first rationalization of the quite different temperature dependences observed for the slip lengths, in terms of relative values of bulk and surface activation energies of the corresponding bulk and surface dissipation processes.

\textbf{Materials and methods.}
A silanol terminated PDMS melt with number average molecular weight $M_{\mathrm n}=685$~kg\(\cdot\)mol$^{-1}$, \textit{\DJ}~=~1.22 was used. This melt was obtained by controlled fractionation of a commercial batch (ABCR Petrarch PS349.5), and mixed with 1wt$\%$ of fluorescent labeled photobleachable PDMS chains with a number average molecular weight $M_{\mathrm n} = 321$~kg\(\cdot\)mol$^{-1}$, \textit{\DJ}~=~1.18. The chains were labeled at both chain-ends with nitrobenzoxadiazole (NBD) fluorescent groups emitting at 550~nm when excited at 458~nm~\cite{leger1996,cohen_synthesis_2012}. \textcolor{black}{This particular label has been chosen for its good miscibility with PDMS. The 1~\% concentration in labeled chains has been chosen to ensure that the dynamic properties (viscosity, self-diffusion) of the bulk melt are not noticabely affected by the presence of the labeled chains while the large molecular weight of the labeled chains ensures that there is no segregation of the labed chains towards the surface (as revealed by EWILF) \cite{leger1995role}.} The liquid was sandwiched between two surfaces separated by mylar spacers of thickness $h=100~\mu$m. The top solid was made of fused silica and was cleaned with a piranha solution~\cite{kern_1970} just before assembling the flow cell. The bottom surface, on which slip was investigated, was the polished surface of a 3~mm thick silicon wafers having either a covalently grafted SAM of OTS~\cite{silberzan_silanation_1991} or PDMS brushes of molecular weight 2~kg$\cdot$mol$^{-1}$ covalently grafted by hydrosylilation~\cite{marzolin_2001}. The fabrication procedures are detailed in the supplementary materials. The contact angle of water on these surfaces was close to 115~\degree ~and the advancing contact angle of dodecane was $\theta_\mathrm{a}=34$~\degree ~with an hysteresis of 1~\degree. 

The experimental technique used to measure the slip lengths is described in detail in the supplementary materials and with a discussion on the resolution in~\cite{henot_macromol_2017}. It is an improved version of the velocimetry technique described by L\'eger {\em et al.}~\cite{leger_wall_1997}. As can be seen in Figure~\ref{schema}a, the determination of the slip length relies on the observation of the evolution under simple shear of a pattern drawn in the fluorescent polymer using photobleaching. The pattern is a line which is vertical in one direction and tilted in the direction perpendicular to the shear. Because of this simple geometry, the observation of the pattern from the top allows one to follow independently portions of the liquid as a function of their vertical position between the two plates. Hence the displacement field of the fluid under shear can be reconstructed. This allows for independent measurements of the slip lengths at both surfaces and of the real shear rate $\dot{\gamma}$ experienced by the polymer melt during the shear. The whole experiment is mounted in a temperature controlled box. 

The solid-friction measurements were performed on an evolution of the apparatus described by Bureau~\textit{et al.}~\cite{bureau_sliding_2004} and later by Cohen~\textit{et al.}~\cite{cohen_incidence_2011}. It consists in a millimetric semi-spherical lens made of cross-linked PDMS in contact with a planar surface. The lens can move at a chosen velocity $V$ and both the friction force $F$ and the contact area $A$ are monitored during the experiment. The friction force is obtained by monitoring the deflection of a double beam cantilever.

\textbf{Results.}
\begin{figure}[htbp]
  \centering
  \includegraphics[width=8.5cm]{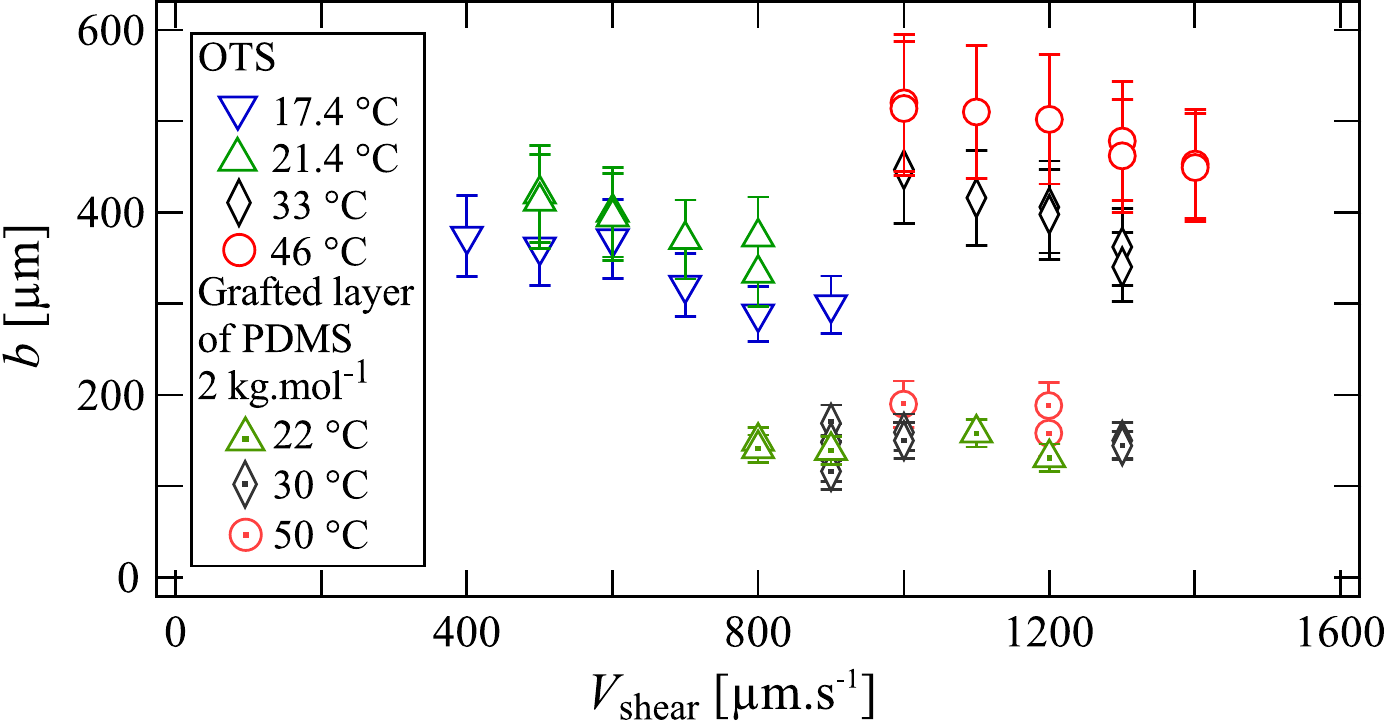}
 \caption{Slip length of a PDMS melt of molecular weight 685~kg$\cdot$mol$^{-1}$ as a function of the shear velocity for different temperatures on a OTS surface and on a grafted layer of PDMS.}
  \label{b_vs_Vshear}
\end{figure}
The slip lengths of a PDMS melt investigated at temperatures going from 17.4~\degree C to 50~\degree C  on an OTS surface and on a grafted layer of PDMS are reported in Figure~\ref{b_vs_Vshear} as a function of velocity of the top plate $V_\mathrm{shear}$. The range of investigated velocities $V_\mathrm{shear}$ is fixed by the setup limits for high velocities and by a small amount of adsorbed chains of the melt on the surfaces for small velocities, leading to a slip transition \cite{leger_wall_1997}. First, we see a significant difference in slip length between the two surfaces as previously observed on the same system~\cite{henot_macromol_2017}. On OTS, on which the slip lengths are larger, they clearly depend on the temperature. We observe a  40 \% increase in slip length when the temperature is increased from 17.4~\degree C to 46~\degree C. For each temperature, the observed slow decrease of the slip lengths when increasing shear rate can be related to the shear thinning of the melts for the shear rates of the experiments ~\cite{henot_acs_2018}. In contrast, on the grafted layer of PDMS, the temperature dependence is significantly weaker.

\begin{figure}[htbp]
  \centering
  \includegraphics[width=8.5cm]{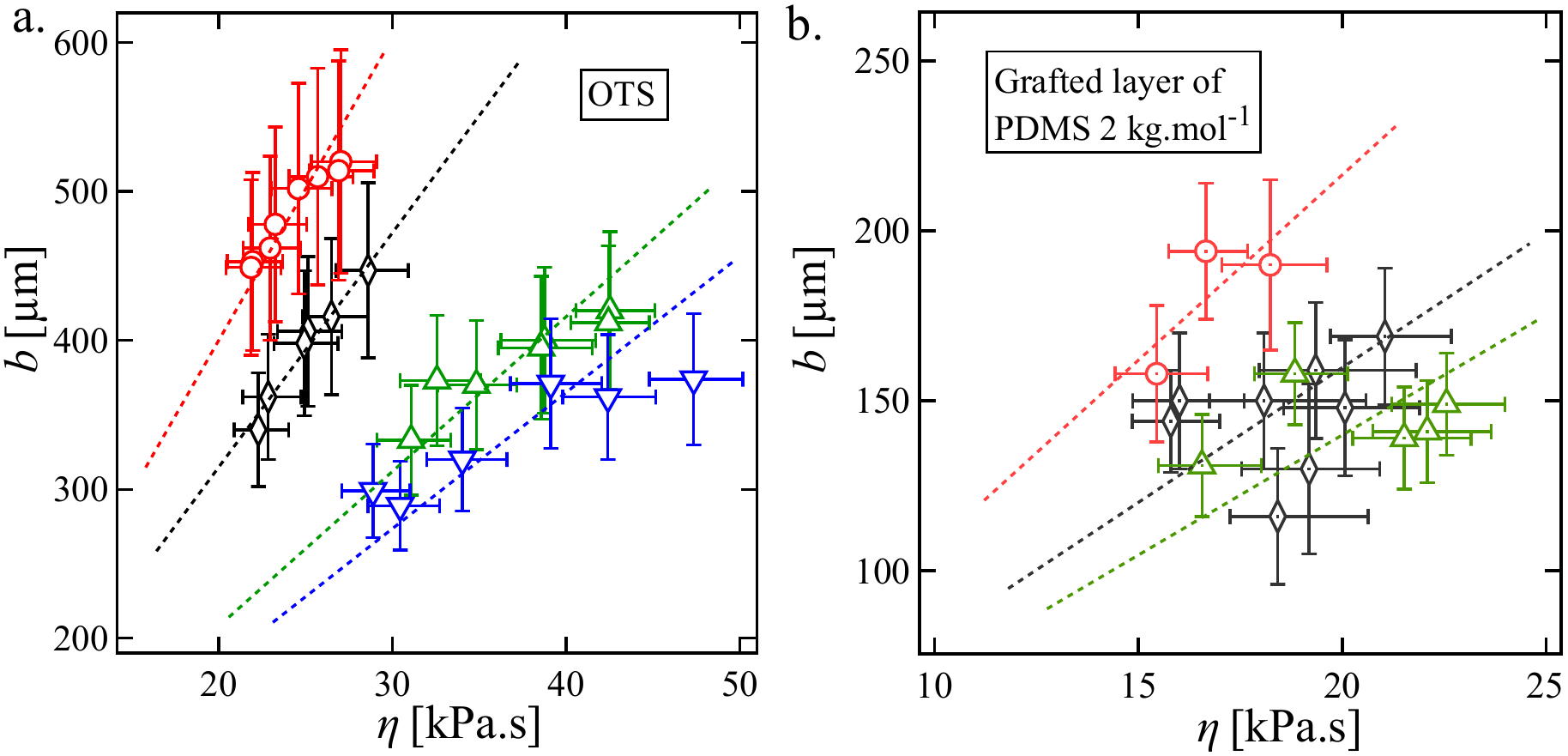}
 \caption{Slip length of a PDMS melt of molecular weight 685~kg$\cdot$mol$^{-1}$ as a function of the viscosity at the given shear rate for different temperatures on an OTS surface (a) and on a grafted layer of PDMS (b). The dashed lines are linear adjustments of the data. The legend of these markers is the same as in Figure~\ref{b_vs_Vshear}.}
  \label{b_vs_eta}
\end{figure}
To be more quantitative, the rheological properties of the melt were measured (see Supplementary materials. Knowing the shear velocities and the slip length, we can deduce the real shear rate experienced by the fluid, and plot in Figure~\ref{b_vs_eta}, the slip lengths as a function of the viscosity at the corresponding shear rates. For each temperature, a linear relation between $b$ and $\eta$ is observed. The slope, which corresponds to the inverse of the interfacial or Navier's friction coefficient $k_\mathrm{melt}$, appears to depend both on the chemical nature of the surface and on the temperature. We see  that, on both surfaces, this friction coefficient decreases with the temperature.
%In figure~\ref{b_vs_eta}, the slip length can be varied on a large range by changing the temperature and the shear rate. For example at $\eta = 15.3$~kPa$\cdot$s, $b(T = 33$~\degree C, $\dot{\gamma} = 2.9$~s$^{-1}) = 340~\mu$m and $b(T = 46$~\degree C, $\dot{\gamma} = 1.8$~s$^{-1}) = 510~\mu$m. This is consistent with what was observed by B\"aumchen~\textit{et al.}~\cite{baumchen_reduced_2009,baumchen_slippage_2012} by varying the temperature and the molecular weight of PS melts.

\begin{figure}[htbp]
  \centering
  \includegraphics[width=8.5cm]{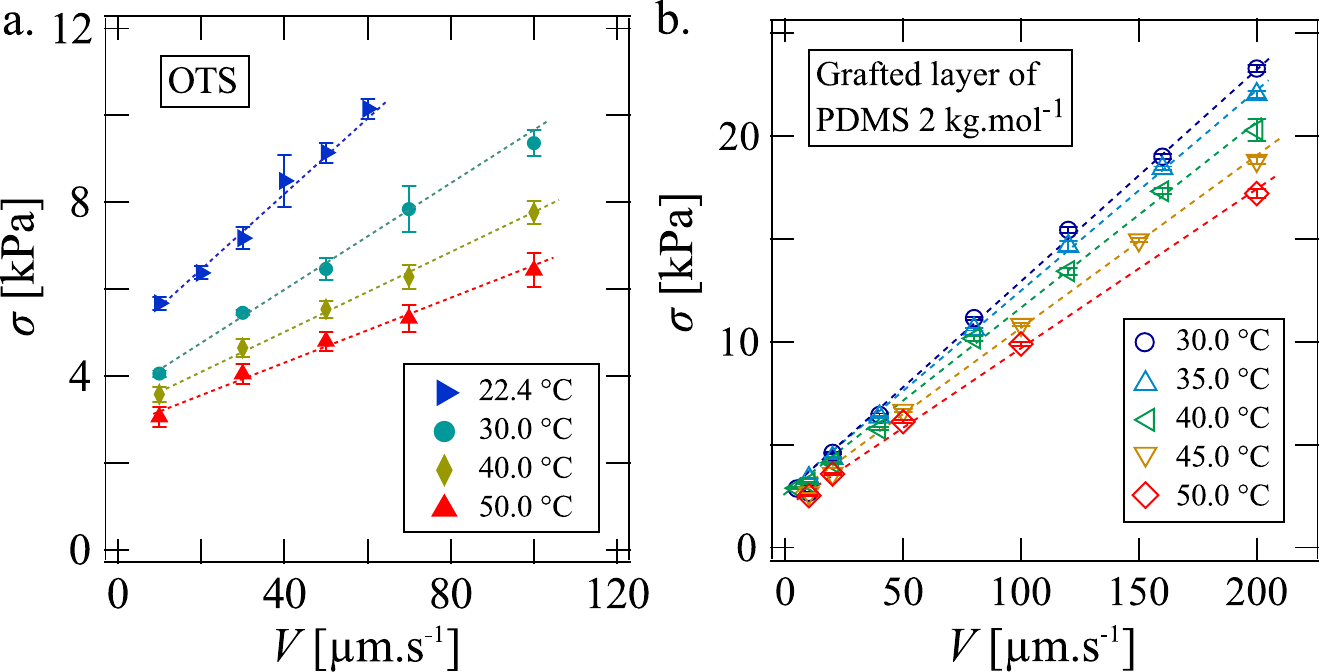}
 \caption{Friction stress of a cross-linked lens of PDMS on an OTS surface (a) and on a grafted layer of PDMS (b) as a function of the sliding velocity. The measurements were done for different temperatures. The dashed lines represent linear fits of the data.}
  \label{contrainte_vs_V}
\end{figure}
In order to gain a better insight into the friction mechanisms, we have also measured the solid sliding friction of a cross-linked PDMS elastomer lens on the same surfaces as a function of the temperature. From the friction force $F$ and the contact area $A$ directly accessible in the experiment as a function of the sliding velocity $V$, we deduced the friction stress $\sigma=F/A$.
Figure~\ref{contrainte_vs_V} shows the friction stress as a function of the velocity for different temperatures on both surfaces. The sliding velocity is limited to 100~$\mu$m$\cdot$s$^{-1}$ on OTS and 200~$\mu$m$\cdot$s$^{-1}$ on the grafted PDMS layers, due to contact instabilities related to the apparition of Schallamach waves at the rubber-surface interface~\cite{schallamach_1971}. At a given temperature, the velocity dependence is linear, of the form $\sigma(V)=\sigma_0+k_\mathrm{elastomer}V$ with the constants $\sigma_0$ and $k_\mathrm{elastomer}$ depending on temperature and on the chemical nature of the surface. For the grafted layer of PDMS, the behavior is compatible with that observed by Bureau and Cohen~\cite{bureau_sliding_2004,cohen_incidence_2011} on the same surface at room temperature.
It can be seen that the values of the friction coefficients are almost twice lower on OTS than on PDMS layers.  It can be noticed that our results are close in order of magnitude to those of Vorvolakos~\textit{et al.}~\cite{vorvolakos_effects_2003} on hexadecylsiloxane SAM who measured $\sigma(65~\upmu$m$\cdot$s$^{-1}) = 11$~kPa at 45~\degree C. This value falls between our OTS and grafted PDMS layer data.

\begin{figure}[htbp]
  \centering
  \includegraphics[width=6cm]{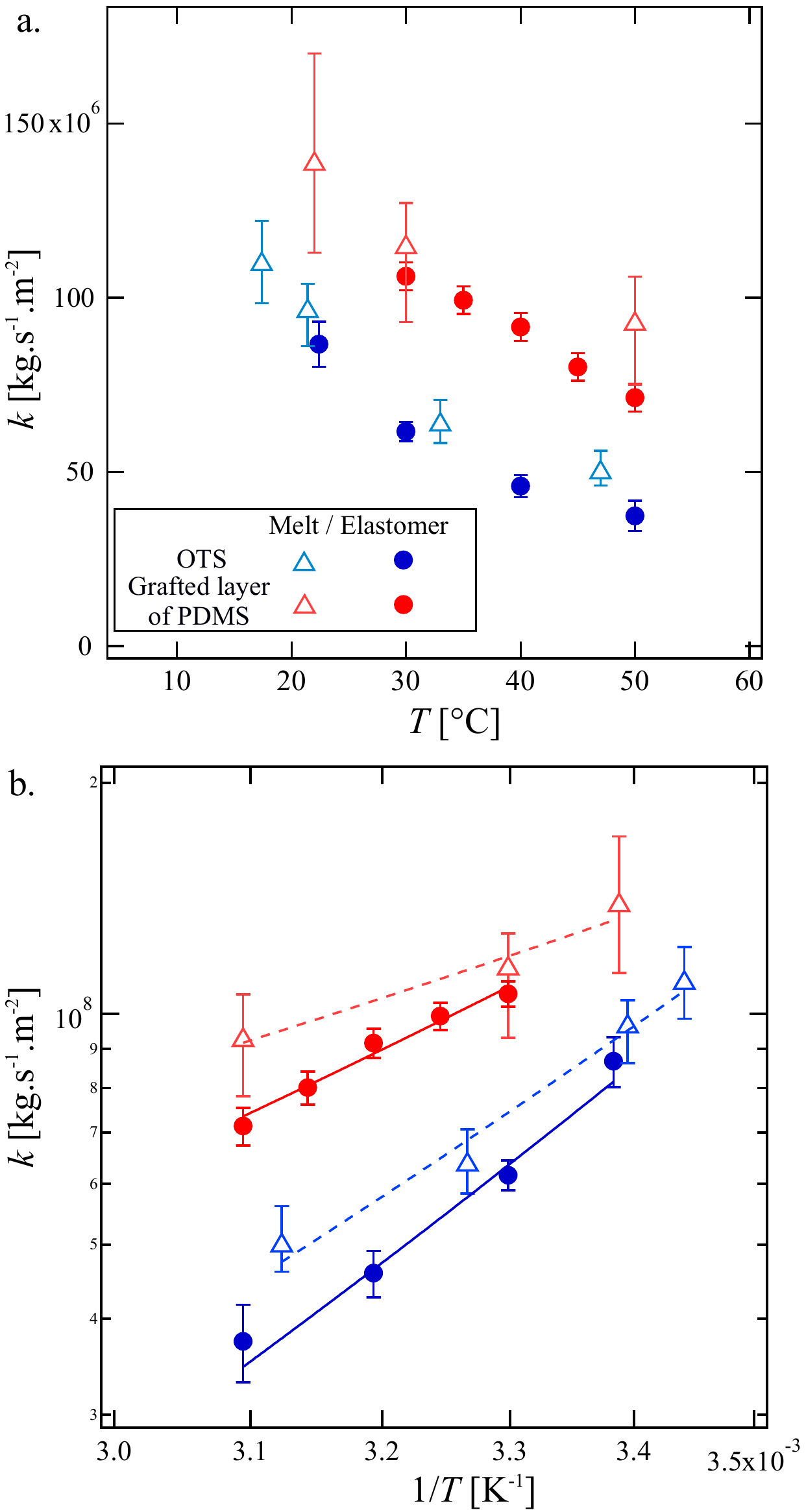}
 \caption{(a) Friction coefficient of a cross-linked lens of PDMS (solid markers) and of a PDMS melt (empty markers) on an OTS surface (blue markers) and on a grafted PDMS layer (red markers) as a function of temperature. (b) The friction coefficient for the same data plotted in a logarithmic scale as a function of the inverse of the temperature. The red and blue curves represent \textcolor{black}{the fits of an activated processes.}}
  \label{k_vs_T}
\end{figure}

Figure~\ref{k_vs_T}.a gathers both solid friction coefficient $k_\mathrm{elastomer}$ and interfacial Navier's coefficient $k_\mathrm{melt}$ as a function of the temperature on respectively the OTS SAM and the grafted PDMS layer. All friction coefficients depend on the chemical nature of the surface. It appears that the solid friction coefficient of the PDMS elastomer and the interfacial friction coefficient of the PDMS melt on the same surface are equal. Note there is no adjustable parameter in this comparison. The friction coefficients decrease with increasing temperature, with a faster decrease on the OTS surface than on the grafted PDMS layer.

Indeed, the temperature dependence of the friction of PDMS elastomers on solids  was investigated by Schallamach~\cite{schallamach_1953,schallamach_1971}, Grosch~\cite{grosch_1963} and later by Ronsin~\textit{et al.}~\cite{Ronsin_2001} and Vorvolakos~\textit{et al.}~\cite{vorvolakos_effects_2003}. They reported a decrease of the friction stress with increasing temperature, as observed here. They invoked an activated attachment-detachment mechanism for the solid friction, so that the friction obeys an Arrhenius law with an activation energy $E_\mathrm{elastomer}$: $k_\mathrm{elastomer}\propto\exp(E_\mathrm{friction}/RT)$, with $R$ the ideal gas constant. Following similar assumptions, the solid friction and the interfacial Navier's coefficients are reported as a function of the  inverse of the temperature in a log-lin scale plot, in Figure~\ref{k_vs_T}b. \textcolor{black}{On the OTS surface and on the grafted layer of PDMS, we measure activation energies of $E_\mathrm{friction,\,melt}(\mathrm{OTS}) = 21.3\pm 4.0 $~kJ$\cdot$mol$^{-1}$, $E_\mathrm{friction,\,elastomer}(\mathrm{OTS}) = 24.6\pm 3.3 $~kJ$\cdot$mol$^{-1}$, $E_\mathrm{friction,\,melt}(\mathrm{PDMS}) = 10.6\pm 7.1$~kJ$\cdot$mol$^{-1}$ and $E_\mathrm{friction,\,elastomer}(\mathrm{PDMS}) = 15.8\pm2.3$~kJ$\cdot$mol$^{-1}$, respectively. The error bars have been estimated with an error-weighted least-squares fit to ln(k) vs. 1/T. The activation energies found by the two different experiments are consistent within the error bars.}  For comparison, Vorvolakos measured an activation energy of $E_\mathrm{friction} = 25$~kJ$\cdot$mol$^{-1}$ for PDMS on a monolayer of grafted hexadecylsiloxane~\cite{vorvolakos_effects_2003}.

The bulk viscosity of a polymer melt is also considered as an activated process in the Newtonian regime (for temperature much higher than the glass transition temperature. Here $T_g=-127$~\degree C.). It is thus common to write $\eta\propto\exp(E_\mathrm{viscous}/RT)$. \textcolor{black}{Using the WLF theory on our experiments (see Supplementary material citing \cite{cox1958,amouroux_effect_2003,chaudhury_1991,chenneviere_2013}), an activation energy for a  PDMS viscous flow was calculated: $E_\mathrm{viscous}=16.3\pm2.8$~kJ$\cdot$mol$^{-1}$, in good agreement with the literature ($E_\mathrm{viscous} =15$~kJ$\cdot$mol$^{-1}$)~\cite{barlow_1964, kataoka_1966}.}

As the slip length is given by the ratio of the viscosity by the interfacial friction coefficient, the temperature dependence of the slip length simply results from the comparison between the bulk molecular movement activation energy $E_\mathrm{viscous}$ and the surface molecular movement activation energy $E_\mathrm{friction}$:
\begin{equation}
b(T) \propto \exp \left(\frac{E_\mathrm{viscous}-E_\mathrm{friction}}{RT}\right)
\end{equation}
On the OTS surface, $E_\mathrm{viscous}<E_\mathrm{friction}$, leading to increasing slip lengths with temperature. On the grafted PDMS layer, $E_\mathrm{viscous}\approx E_\mathrm{friction}$, which leads to a weak temperature dependence, as observed in our experiments.

\textcolor{black}{The proposed mechanism of thermodynamically activated friction leading to eq.(\ref{equ_b}) predicts also that the slip length can decrease with the temperature if $E_\mathrm{viscous}>E_\mathrm{friction}$. This is not observed in the two cases of this article, but B\"aumchen~\textit{et al.} reported for polystyrene (13.7~kg$\cdot$mol$^{-1}$) flowing on DTS a slip length decreasing from 6 $\upmu$m to 1 $\upmu$m when the temperature is increased from 383 K to 403 K \cite{baumchen_sliding_2010}. It should be noticed however, that these experiments have been performed close to the glass transition temperature where the use of a simple activation mechanism could be questionable both for $\eta(T)$ and $k(T)$. As a consequence, our model is probably too simple to be generalized to glassy system even though it gives a good qualitative explanation of the observed temperature dependences.}

% rheofludifiant (à voir)

% This is not expected to be a general behavior and depending on the polymer melt or on the surface, other monotony could be possible. For instance, in the case of a PDMS melt slipping on a grafted layer of PDMS,  we can expect both activation energies to be equal and thus a slip length independent of the temperature.
% Baumchen : constant ou decroissant

In conclusion, we have measured the temperature dependence of the hydrodynamics boundary condition between a PDMS polymer melt and two different surfaces allowing one to extract the friction stress exerted by the fluid on the solid surface. We characterized independently the temperature dependence of a cross-linked PDMS elastomer on the same surfaces, thus confirming the identity of the solid and of the Navier's friction coefficients. We showed that both the friction stress of a liquid and of an elastomer decrease with the temperature following Arrhenius laws. We conclude that either an increase or a decrease of the slip length with respect to the temperature can be observed, depending on the compared values of the bulk viscosity and of the interfacial friction activation energies. This new result shines some light on the molecular mechanisms which determine the hydrodynamic boundary condition in polymeric or simple fluids.

{\bf Acknowledgements:} This  work  was  supported  by  ANR-ENCORE program (ANR-15-CE06-005). We thank F. Boulogne, A. Chennevi\`ere and O. B\"aumchen for interesting discussions.

\textcolor{black}{M.H. and M.G. equally contributed to this paper.}
\bibliographystyle{apsrev4-1}
\bibliography{bibliographie}

%\newpage
%Table of content graphics:\\
%\begin{figure}[htbp]
%  \centering
%  \includegraphics[width=8cm]{TOC2.pdf}
%  \label{toc}
% \end{figure}
\end{document}